\documentclass[conference]{IEEEtran}

\usepackage[pdftex]{graphicx}
\graphicspath{{pic/}}

\usepackage[cmex10]{amsmath}
\usepackage[cmintegrals]{newtxmath}
\usepackage{mathrsfs}

\ifCLASSOPTIONcompsoc
 \usepackage[caption=false,font=normalsize,labelfont=sf,textfont=sf]{subfig}
\else
 \usepackage[caption=false,font=footnotesize]{subfig}
\fi

\usepackage{url}

\usepackage{cite}

\usepackage{multirow}
\usepackage{array}
\newcolumntype{L}[1]{>{\raggedright\let\newline\\\arraybackslash\hspace{0pt}}m{#1}}
\newcolumntype{C}[1]{>{\centering\let\newline\\\arraybackslash\hspace{0pt}}m{#1}}
\newcolumntype{R}[1]{>{\raggedleft\let\newline\\\arraybackslash\hspace{0pt}}m{#1}}

\usepackage{siunitx}

\newcommand{\erfc}{\mathrm{erfc}}
\newcommand{\binomial}[2]{\mathcal{B}\left( #1 , #2 \right)}

\usepackage{todonotes}
\definecolor{morange}{rgb}{0.8,0.2,0}
\definecolor{mblue}{rgb}{0,0.3,1.0}
\definecolor{mgreen}{rgb}{0.2,0.4,0}

\newcommand{\rxi}[1]{\text{Rx}_{#1}}
\newcommand{\txi}[1]{\text{Tx}_{#1}}
\newcommand{\ntx}{N_\text{Tx}}

\newcommand{\txt}{2\!\times\! 2}

\newcommand{\Ft}[1]{F(#1)}

\newcommand{\Fij}[1]{F_\text{#1}}

\newcommand{\uvec}{\mathbf{u}}

\newcommand{\uik}[2]{u_{#1}[#2]}

\newcommand{\xik}[2]{x_{#1}[#2]}

\newcommand{\svec}{\mathbf{s}}

\newcommand{\sk}[1]{s_{#1}}
\newcommand{\yk}[1]{y[#1]}
\newcommand{\yik}[2]{y_{#1}[#2]}
\newcommand{\ycombk}[2]{y_\text{#1}[#2]}
\newcommand{\nk}[1]{n[#1]}
\newcommand{\nik}[2]{n_{#1}[#2]}
\newcommand{\uhatk}[1]{\hat{u}[#1]}
\newcommand{\utildek}[1]{\tilde{u}[#1]}

\newcommand{\hjil}[3]{h_{#1#2}[#3]}

\newcommand{\hl}[1]{h[#1]}
\newcommand{\hhatl}[1]{\hat{h}[#1]}

\newcommand{\G}{\mathbf{G}}
\newcommand{\I}{\mathbf{I}}
\newcommand{\hermit}{\mathrm{H}}

\newcommand{\Ts}{T_\text{s}}

\newcommand{\xth}[1]{{#1\text{th}}}

\begin{document}
\title{Spatial Coding Techniques for Molecular MIMO}

\author{\IEEEauthorblockN{Martin~Damrath\IEEEauthorrefmark{1},
H.~Birkan~Yilmaz\IEEEauthorrefmark{2},
Chan-Byoung~Chae\IEEEauthorrefmark{2}, and 
Peter~Adam~Hoeher\IEEEauthorrefmark{1}}
\IEEEauthorblockA{\IEEEauthorrefmark{1}Faculty of Engineering, Kiel University, Germany, Email: \{md, ph\}@tf.uni-kiel.de}
\IEEEauthorblockA{\IEEEauthorrefmark{2}School of Integrated Technology, Yonsei University, Korea, Email: \{birkan.yilmaz, cbchae\}@yonsei.ac.kr}}

\maketitle

\begin{abstract}
This paper studies spatial diversity techniques applied to multiple-input multiple-output (MIMO) diffusion-based molecular communications (DBMC).
Two types of spatial coding techniques, namely Alamouti-type coding and repetition MIMO coding are suggested and analyzed.
In addition, we consider receiver-side equal-gain combining, which is equivalent to maximum-ratio combining in symmetrical scenarios.
For numerical analysis, the channel impulse responses of a symmetrical $\txt$ MIMO-DBMC system are acquired by a trained artificial neural network.
It is demonstrated that spatial diversity has the potential to improve the system performance and that repetition MIMO coding outperforms Alamouti-type coding.\\
\end{abstract}


\IEEEpeerreviewmaketitle

\section{Introduction}
Molecular communication (MC) is a biologically inspired communication paradigm,
where molecules are the information carriers \cite{farsad2016comprehensiveSO}.
MC is claimed to be a key technology in realizing autonomous nanomachines (NMs), which size ranges from several nanometers up to a few micrometers \cite{Xia2003}.
The capability of NMs can be enhanced by working in a cooperative manner \cite{guo2016molecularCC,Nakano2013}. Therefore, communication at small scales is at a crucial point.
The main application is anticipated to be in the medical sector,
where NMs can be used for applications like targeted drug delivery, tissue engineering, or health monitoring \cite{Nakano2013}.

In diffusion-based molecular communication (DBMC) \cite{Pierobon2010},
messenger molecules propagate, according to the law of diffusion, from a source to the sink.
While this propagation is energy efficient,
this communication channel is fundamentally different from the classical radio-based wireless communication channel.
For DBMC, the channel impulse response is slowly decreasing, which causes intersymbol interference (ISI) and unreliable transmission~\cite{noel2014optimalRD}. 
In order to improve this unreliable transmission, spatial diversity can be exploited in multiple-input multiple-output (MIMO) scenarios with multiple antennas at the transmitter and/or receiver side.

MIMO is a familiar topic in classical wireless communication.
In molecular communication, however, it has rarely been considered.
The authors in \cite{Meng2012a} were the first to study molecular communication in conjunction with MIMO.
They proposed different techniques for transmitter diversity, diversity combining at the receiver side, and spatial multiplexing.
Their focus, however, was on multi-user interference, thus neglecting ISI throughout the work.
In \cite{Koo2016}, the authors modeled a MIMO channel taking into account ISI and interlink interference (ILI).
They focused then on spatial multiplexing and proposed different detection algorithms.
The authors applied their algorithms to a tabletop molecular MIMO testbed and demonstrated an improvement in the data rate compared to their single-input single-output (SISO) case.
In \cite{Lu2016}, the authors analyzed the influence of a second absorbing receiver on bit error ratio (BER) and capacity for a broadcast MC system.

In this work, the focus is on a MIMO channel considering ISI and ILI.
The main contribution is the analysis of different spatial diversity algorithms at the transmitter side. For the transmitter side, we propose Alamouti-type coding and repetition MIMO coding;  for the receiver side, we propose equal-gain combining that is equivalent to maximum-ratio combining in symmetrical scenarios.
The diversity gain compared to a SISO scenario is investigated by means of a BER simulation,
where the influence of the system parameters is shown.
The MIMO channel impulse responses are acquired by a trained artificial neural network (ANN).



\IEEEpubidadjcol

\section{System Model\label{Sec:SystemModel}}

\subsection{Topology and Propagation Model}
\begin{figure} 
 \centering 
 \def\svgwidth{\columnwidth-1cm}
  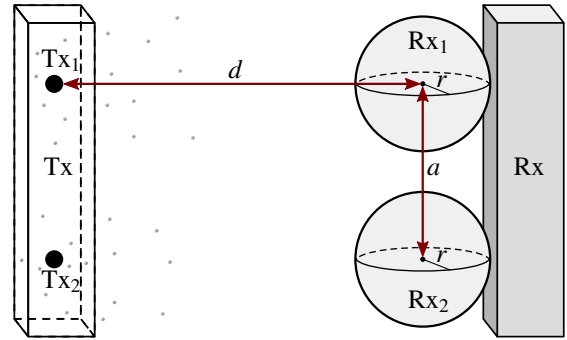
 \caption{Model of the diffusion-based molecular $2\times2$ MIMO system \cite{Koo2016}.}
 \label{Fig:SystemModel}
\end{figure}
The system model under investigation is similar to the system model introduced in \cite{Koo2016} and \cite{lee2015molecularMC}.
As shown in Fig.~\ref{Fig:SystemModel}, it consists of a transmitter Tx and a receiver Rx in an infinite three-dimensional homogeneous fluid medium without drift.
In the sense of a $\txt$~MIMO system, the Tx includes two point antennas $\txi{1}$ and $\txi{2}$,
while the Rx includes two spherical receive antennas $\rxi{1}$ and $\rxi{2}$ with radius $r$ that are attached to the reflecting body of the Rx.
Throughout this work, a symmetrical scenario is assumed where $\txi{1}$ is aligned to $\rxi{1}$ and $\txi{2}$ is aligned to $\rxi{2}$.
As a result, the distance between $\txi{1}$ and $\rxi{1}$, as well as between $\txi{2}$ and $\rxi{2}$, is given as $d$.
Furthermore, the separation distance between $\txi{1}$ and $\txi{2}$, as well as between $\rxi{1}$ and $\rxi{2}$, is given as $a$.
The fluid medium is described by the diffusion coefficient $D$.

The molecules emitted by $\txi{1}$ and $\txi{2}$ propagate by Brownian motion, which is described by the Wiener process~\cite{srinivas2012molecular}. Whenever a diffusing molecule hits $\rxi{1}$ or $\rxi{2}$,
it will be counted and perfectly absorbed, i.e., it will be removed from the environment.
As a result, the time histogram of absorbed molecules at $\rxi{1}$ and $\rxi{2}$ follow the first passage time concept.
For a SISO scenario in a 3-dimensional (3-D) environment, there exists a closed-form formula which describes the probability that a molecule hits Rx until time $t$ after its release \cite{Yilmaz2014}:
\begin{equation}\label{Eq:Fhit}
 \Ft{t} = \frac{r}{d} \ \erfc \left( \frac{d-r}{\sqrt{4Dt}} \right),
\end{equation}
where $\erfc(\cdot)$ is the complementary error function.
However, for multiple absorbing spheres inside the medium, no such closed-form expression exists. 
Thus, a corresponding expression to (\ref{Eq:Fhit}) for a given MIMO scenario has to be obtained by either random-walk-based simulations or by using a trained ANN as presented in Section~\ref{Sec:ANN}.

\subsection{Communication Channel}
The modulation scheme under investigation is on-off keying (OOK)~\cite{kuran2011modulationTF,yilmaz2014simulationSO,kim2013novel}. 
$\txi{i}$ emits either no molecules or $N$ messenger molecules at the beginning of a symbol period of length $\Ts$ to represent bit $\uik{i}{k}=0$ or $\uik{i}{k}=1$, respectively.
Molecules emitted by $\txi{1}$ and $\txi{2}$ are of the same type.
Furthermore, Rx is assumed to be synchronized with Tx in time domain as suggested in \cite{Moore2013}.
In addition, $\rxi{1}$ and $\rxi{2}$ perform strength\slash energy detection at each symbol duration \cite{Mahfuz2010b,Llatser2013}.

The MIMO channel can be separated into subchannels from each transmit antenna $\txi{i}$ to each receive antenna $\rxi{j}$.
Each subchannel is thereby characterized by the corresponding channel coefficients $\hjil{j}{i}{\ell}$ ($0 \leq \ell \leq L$),
which describe the probability that a molecule hits $\rxi{j}$ during the $\ell$th time slot after its emission at $\txi{i}$.
All subchannels can be represented by an equivalent discrete-time channel model with effective channel memory length $L$ \cite{Damrath2017,genc2016isiAM}.
As a result, the number of received molecules at $\rxi{j}$ can be described by the summation over all subchannels related to $\rxi{j}$ as 
\begin{equation}
  \yik{j}{k} = \sum \limits_{i=1}^{\ntx} \sum \limits_{\ell=0}^L \hjil{j}{i}{\ell} \xik{i}{k\!-\!\ell} + \nik{j}{k},
\end{equation}
where $\ntx$ is the number of transmitters, $\nik{j}{k}$ describes the amplitude dependent noise caused by the diffusive propagation of the molecules,
and $\xik{i}{k}$ is the discrete-time representation of the modulated data symbol transmitted by $\txi{i}$ at the start of the $\xth{k}$ transmission interval.
For OOK it is defined as
 \begin{equation} \label{Eq:Mapping}
  \xik{i}{k}  = \begin{cases}
            N &\quad \textnormal{if} \ \uik{i}{k}=1 \\
            0 &\quad \textnormal{if} \ \uik{i}{k}=0.
           \end{cases}
 \end{equation}
Since the hitting process of molecules during a bit period can be described by a binomial distribution~\cite{yilmaz2014arrivalMF},
$\yik{j}{k}$ is represented by the sum over binomial distributions
\begin{equation}
 \yik{j}{k} \sim \sum \limits_{i=1}^{\ntx} \sum \limits_{\ell=0}^L \binomial{\xik{i}{k\!-\!\ell}}{\hjil{j}{i}{\ell}},
\end{equation}
where $\binomial{M}{p}$ describe a binomial distribution with $M$ number of trials and success probability $p$.

With the help of (\ref{Eq:Fhit}), the channel coefficients for a SISO scenario can be easily calculated by
\begin{equation}\label{Eq:hSISO}
 \hl{\ell} = \Ft{(\ell+1)\Ts} - \Ft{\ell\Ts}.
\end{equation}
However, for multiple absorbing spheres inside the medium,
the channel coefficients $\hjil{j}{i}{\ell}$ for a given MIMO scenario has to be obtained by either random-walk-based simulations or by using a trained ANN as presented in Sec.~\ref{Sec:ANN}.

\section{ANN for Channel Modeling\label{Sec:ANN}}
For modeling a molecular MIMO channel, we utilized the trained ANN of our previous work~\cite{lee2017machineLB_ARXIV}.
A trained ANN is able to estimate the channel coefficients $\hjil{j}{i}{\ell}$ for a given MIMO scenario without running simulations.

In a first step, we defined an expected analytical channel response function by introducing fitting parameters into (\ref{Eq:Fhit}).
The analytical channel response function at $\rxi{1}$ is defined as follows:
\begin{align}
\begin{split}
\Fij{11} (t, b_1, b_2, b_3) =  b_1 \, \frac{r}{d} \,\erfc \left( \frac{d\!-\!r}{(4D)^{b_2} \, t^{b_3}}\right),
\end{split}
\label{eqn_model_rx1}
\end{align}
where $b_1$, $b_2$, and $b_3$ represent the model fitting parameters.
Similarly we define the response at $\rxi{2}$ (due to the cross link interference) as follows:
\begin{align}
 \Fij{21} (t, b_4, b_5, b_6) =  b_4 \, \frac{r}{\sqrt{d^2\!+\!a^2}} \,\erfc \left( \frac{\sqrt{d^2\!+\!a^2}-r}{(4D)^{b_5} \, t^{b_6}}\right),
\end{align}
where $b_4$, $b_5$, and $b_6$ are also model fitting parameters.

In a second step, we fitted the expected analytical channel response functions to data obtained in extensive simulations.
To determine the $b_i$ values, we use a nonlinear least squares curve-fitting technique.
These values, in conjunction with selected reference system parameters $d$, $a$, $r$, and $D$, are the basis of training and test datasets.
Hence, the output of the curve-fitting process consists of the model parameters $b_i$ for each selected simulation scenario.

In a third step, after forming the training and test datasets, the training data is fed to the ANN training process.
Note that the trained ANN does not require any simulation data.
That is, the required inputs are arbitrary system parameters $d$, $a$, $r$, and $D$.
After training, the ANN is able to predict the fitting parameters $b_i$ for these arbitrary system parameters.

\begin{figure}[!t]
	\begin{center}
    	\includegraphics[width=1.0\columnwidth,keepaspectratio]%
		{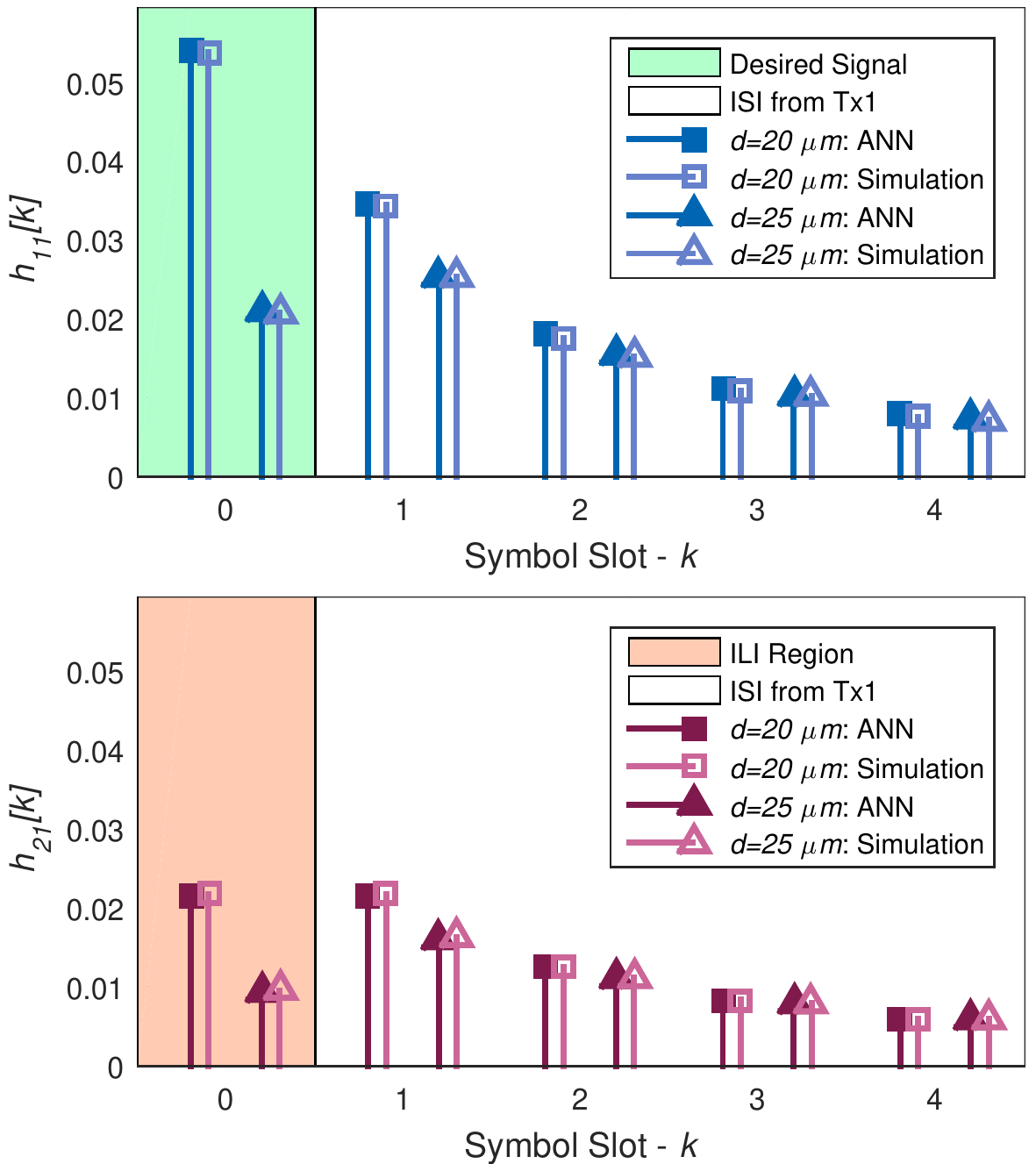}
    \end{center}
	\caption{Comparison of channel coefficients from ANN and simulation data for different distances ($a\!=\!\SI{13}{\micro\meter}$, $r\!=\!\SI{5}{\micro\meter}$, $D\!=\!\SI{200}{\micro\meter^2/\second}$, $\Ts = \SI{0.4}{\second}$)}
	\label{fig_ann_vs_sim_res_hijk}
\end{figure}

In Fig.~\ref{fig_ann_vs_sim_res_hijk}, we present the channel coefficients that are acquired from extensive simulations and the trained ANN.
We plot the $\hjil{1}{1}{k}$ and $\hjil{2}{1}{k}$ values by utilizing $\Fij{11}$, $\Fij{21}$, and the symbol duration.
Our results validate and support the using of ANN to obtain the channel coefficients.

\section{Spatial Diversity\label{Sec:SpatialDiversity}}
Usually, spatial coding is performed along multiple transmit antennas,
whereas combining strategies are applied to multiple receive antennas.
In order to achieve a spatial diversity gain at the Tx side,
the same information is transmitted over multiple antennas.
Therefore, before spatial coding over multiple Tx antennas is applied, the binary data sequence $\uvec$ is mapped onto a sequence of data symbols $\svec$.
Thereby, $\sk{k}$ denotes the $k$th data symbol of $\svec$.
In what follows, two spatial coding techniques are presented - Alamouti-type coding and repetition MIMO coding, respectively.
In addition, equal-gain combining, which is equivalent to maximum-ratio combining (MRC) for symmetrical scenarios, is suggested as combining strategy.

\subsection{Alamouti-type Coding}
The Alamouti scheme \cite{Alamouti1998} is a space-time block code widely used in radio-based communication systems for spatial diversity.
In its origin, the Alamouti code can be represented by the $\txt$ transmission matrix
\begin{equation}
  \G = \begin{bmatrix}
                \sk{k} & \sk{k+1} \\ -\sk{k+1}^* & \sk{k}^*
                \end{bmatrix},
\end{equation}
where the columns of the matrix correspond to the transmit antennas and the rows corresponds to two consecutive transmission intervals $[k\Ts \ (k\!+\!1)\Ts]$ and $[(k\!+\!1)\Ts \ (k\!+\!2)\Ts]$, respectively. 
As a result, in the first time slot, $\xik{1}{k}=\sk{k}$ is transmitted via the first transmit antenna and $\xik{2}{k}=\sk{k+1}$ is transmitted simultaneously via the second transmit antenna.
In the second time slot, $\xik{1}{k\!+\!1}=-\sk{k+1}^*$ is transmitted via $\txi{1}$ and $\xik{2}{k\!+\!1}= \sk{k}^*$ is transmitted simultaneously via $\txi{2}$.

Due to the fact that both transmit antennas emit the same information, a spatial diversity gain can be achieved.
The main advantage of the Alamouti scheme is that it is an orthogonal space-time block code, i.e. $\G^\hermit\G=2\I$,
where $\G^\hermit$ is the Hermitian of matrix $\G$ and $\I$ denotes the identity matrix.
Orthogonality simplifies the implementation of a maximum-likelihood detector, because ILI can be canceled completely.
In the case of ISI, however, orthogonality is getting lost and more complex detection algorithms must be applied such as maximum-likelihood sequence estimation.

In the case of molecular communication, the data symbols (amount of emitted molecules) are non-negative and real-valued rather than complex-valued.
For OOK which is considered throughout this work, data bits are mapped onto data symbols $\sk{k}\in\{0,N\}$ following the principle of (\ref{Eq:Mapping}).
Hence, the classical Alamouti code has to be adapted to an Alamouti-type code that avoids the minus sign and the complex conjugate.
As suggested in \cite{Simon2005},
the adaptation can be done by discarding the complex conjugate operation and replacing negative symbols by $\bar{s}_k:=N-\sk{k}$.
As a result, the transmission matrix of the Alamouti-type code is given by
\begin{equation}\label{Eq:a_G}
  \G = \begin{bmatrix}
                       \sk{k} & \sk{k+1} \\ N-\sk{k+1} & \sk{k}
                     \end{bmatrix}.
\end{equation}

\subsection{Repetition MIMO Coding}
A simple alternative to orthogonal Alamouti codes is offered by repetition MIMO \cite{Wilson2005}.
In repetition MIMO the information is distributed over all transmit antennas, 
where each antenna transmit exactly the same data symbol at the same time.
As a result, the transmission matrix for a $\txt$ MIMO scenario is represented by
\begin{equation}
 \G = \begin{bmatrix}
       \sk{k} & \sk{k}
      \end{bmatrix}.
\end{equation}
A big advantage of repetition MIMO is that, even in the presence of ISI, single antenna detection algorithms can be applied.
In addition, the ILI is constructive and contributes to the signal strength.

\subsection{Equal-gain Combining}
Before detection can be performed at Rx, the received signals of each receive antenna have to be combined/selected in a certain way.
Following the equal-gain combining (EGC) algorithm,
the signals of all receive antennas are equally weighted and combined:
\begin{equation}\label{Eq:EGC}
 \ycombk{}{k}=\yik{1}{k}+\yik{2}{k}.
\end{equation}
Note that, for symmetrical scenarios, EGC is equivalent to MRC.
In MRC, the combined signals are weighted by a factor that is proportional to the corresponding channel quality.
For symmetrical scenarios, however, the channels at both receive antennas are the same.
As a result, the channel description for the investigated scenario, can be further simplified.
Considering that $\hjil{1}{1}{\ell}=\hjil{2}{2}{\ell}$ and $\hjil{1}{2}{\ell}=\hjil{2}{1}{\ell}$,
(\ref{Eq:EGC}) can be restated as
\begin{align}\label{Eq:yEGC}
 \ycombk{}{k}=\sum\limits_{\ell=0}^L \hl{\ell} \left( \xik{1}{k\!-\!\ell}+\xik{2}{k\!-\!\ell} \right) + \nk{k},
\end{align}
where $\hl{\ell}\doteq\hjil{1}{1}{\ell}+\hjil{1}{2}{\ell}$ and $\nk{k}\doteq\nik{1}{k}+\nik{2}{k}$.

\section{Detection Algorithms\label{Sec:DetectionAlgorithms}}
For the bit error analysis throughout this paper, 
two different detection algorithms are considered and adopted from \cite{Damrath2016}.
First of all, the low-complexity adaptive threshold detector (ATD) is applied, which works independent of explicit channel knowledge:
\begin{equation}
 \uhatk{k} =  \begin{cases}
          1 & \quad \text{if } \yk{k} > \yk{k\!-\!1}\\
	  0 & \quad \text{if } \yk{k} \leq \yk{k\!-\!1}.\\
  \end{cases}
  \nonumber
\end{equation}
The second algorithm is maximum-likelihood sequence estimation (MLSE) based on the suboptimal squared Euclidean distance branch metric
\begin{equation}
  \gamma(\yk{k}|\left[\utildek{k},\dots,\utildek{k\!-\!L}\right])=\left( \yk{k} - \sum \limits_{\ell=0}^L N \hhatl{\ell} \utildek{k\!-\!\ell} \right)^2.
  \nonumber
\end{equation}
For a single antenna system $\hhatl{\ell}$ is assumed to be equal to the channel coefficients from (\ref{Eq:hSISO}).
For repetition coding with EGC $\hhatl{\ell}=2\hl{\ell}$, 
where $\hl{\ell}$ is defined as in (\ref{Eq:yEGC}).

In the case of Alamouti-type coding, the branch metric can be further adapted.
Since the information of two bits is spread over two consecutive time slots,
the branch metric can be evaluated jointly for both received numbers of molecules.

\section{Numerical Results\label{Sec:NumericalResults}}
\begin{table}
\caption{Simulation parameters used for analysis. The default parameters are in bold face.}
\centering
\begin{tabular}{L{2.2cm} l}
 \hline $\vphantom{\frac{1}{2}}$Parameter & Value\\ \hline
 $D \ [\si{\micro\meter^2/\second}]$ & $\{ 50, \mathbf{100}, 150, 200 \}$ \\ 
 $r \ [\si{\micro\meter}]$ & $5$ \\ 
 $d \ [\si{\micro\meter}]$ & $20$\\
 $a \ [\si{\micro\meter}]$ & $\{ \mathbf{11}, 13, 15, 17\}$ \\ 
 $\Ts \ [\si{\second}]$ & $0.6$\\
 $L$ & $3$ \\
 $N$ & $\{500,\mathbf{1\:\!000},1\:\!500,2\:\!000 \}$ \\ 
 $K$ & $10^6$ \\ 
 $R$ & $1000$ \\
 \hline
\end{tabular}
\label{Tab:Parameter}
\end{table}
For simulative analysis, different parameters settings are considered.
These are summarized in Table~\ref{Tab:Parameter}, 
where $K$ is the bit sequence length for one channel realization and $R$ is the total number of channel realizations.
It is assumed that after $(L+1)\Ts=\SI{2.4}{\second}$ the remaining ISI is negligible.
In order to guarantee a fair comparison between the SISO and the $\txt$ MIMO system by means of transmitting energy, 
the number of released molecules $N$ in the case of the SISO scenario is set twice as large as for the MIMO case.
Furthermore for the SISO scenario, there is just a single transmit and a single receive antenna present in the environment.

The effect of spatial diversity is examined by means of a BER analysis.
We investigate how the BER is impacted by the number of molecules $N$, separation distance $a$, and diffusion coefficient $D$.
This is done by varying one system parameter,
while fixing the other ones to the bold-faced values in Table~\ref{Tab:Parameter}.

\begin{figure*}
\begin{footnotesize}
\begin{center}

\newcommand*{\FigWidth}{6cm}
\newcommand*{\FigHeight}{4.5cm}
\includegraphics{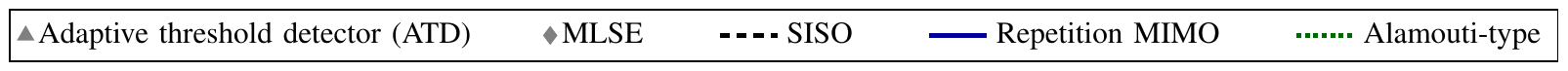}\\
\vspace{-0.3cm}
\subfloat[{Variation of numbers of molecules.}]{
\includegraphics{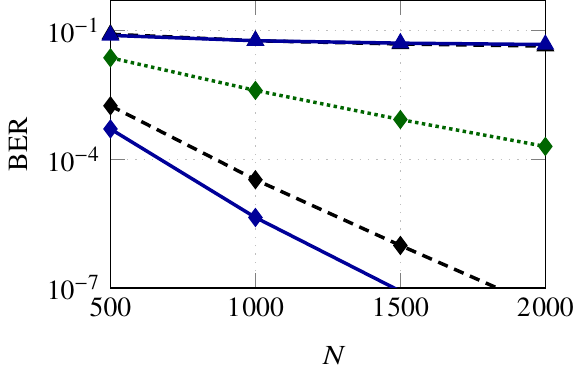}
\label{Fig:BER_Nmol}}
\hfil
\subfloat[{Variation of separation distance.}]{
\includegraphics{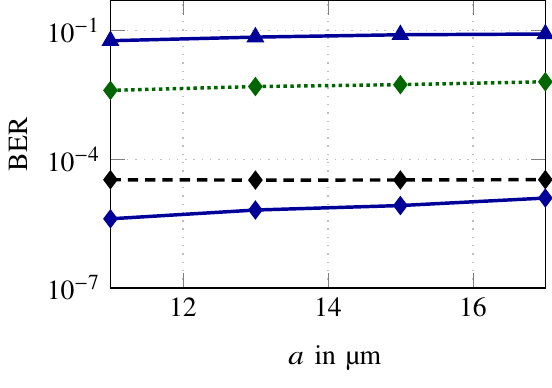}
\label{Fig:BER_a}} 
\hfil
\subfloat[{Variation of diffusion coefficient.}]{
\includegraphics{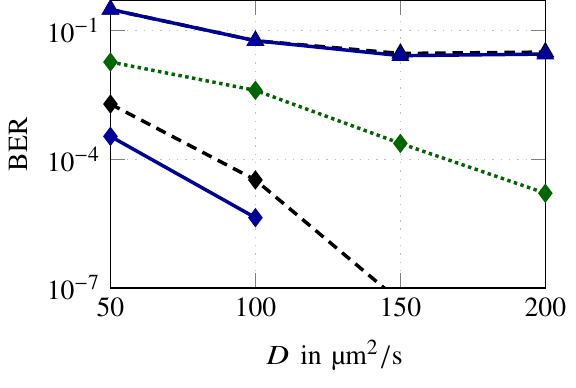}
\label{Fig:BER_D}}
\end{center}

\end{footnotesize}
\caption{Bit error rate performance as a function of the number of molecules \protect\subref{Fig:BER_Nmol},
separation distance \protect\subref{Fig:BER_a}, and diffusion coefficient \protect\subref{Fig:BER_D}.
If the corresponding parameter is not varying, it is fixed to $N=1\:\!000$, $a=\SI{11}{\micro\meter}$, and $D=\SI{100}{\micro\meter^2/\second}$.}
\label{Fig:BER}
\end{figure*}

\emph{Effect of Number of Emitted Molecules:\label{Sec:varN}}
In Fig.~\ref{Fig:BER_Nmol}, the effect of variation on the number of emitted molecules is shown.
The number of molecules is proportional to the signal strength.
In fact, if $N$ is increased, the variance lessens around the expected channel impulse response and more molecules hit the Rx sphere.
Consequently, when $N$ is increased, all detection algorithms achieve better performance when $N$ is increased.
ATD benefits from the ISI in the system \cite{Damrath2016}.
The best BER performance, however, is achieved by MLSE.
This is due to the fact that MLSE implies channel equalization, which counteracts ISI.
Repetition MIMO with ATD slightly outperforms SISO transmission in a region with few molecules.
However, the spatial diversity gain for ATD is not significant.
For MLSE, the spatial diversity gain can be more clearly observed.
The maximum BER improvement of repetition MIMO over the SISO case is by a factor of almost $10$.
A factor of almost $400$ is achieved, 
when power normalization (the emitted number of molecules in the SISO case is twice as large as in the $\txt$ MIMO case) is neglected.
For the system considered here, Alamouti-type coding brings no BER improvement.
The reason is that in case of repetition MIMO the ILI constructively contributes to the signal strength,
whereas in Alamouti-type coding ILI is competitive and as a result more destructive.

\emph{Effect of Antenna Separation:}
Fig.~\ref{Fig:BER_a} shows the effect of the antenna separation distance on the MIMO system performance.
Thereby, all MIMO systems perform worse if the antenna separation distance is increased,
because the spatial gain from ILI is decreased.
For repetition MIMO with MLSE, a spatial diversity gain is still present for $a=\SI{17}{\micro\meter}$.

\emph{Effect of Diffusion Coefficient:}
Fig~\ref{Fig:BER_D} shows the effect of variation in the diffusion coefficient on the system performance.
A larger diffusion coefficient leads to a more spiky channel impulse response. 
As a result, all detection algorithms under consideration achieve a smaller BER if the diffusion coefficient is increased.
The spatial diversity gain from repetition MIMO with ATD increases with $D$.
However, the spatial diversity gain is not significant.
The gap between repetition MIMO with MLSE and the SISO MLSE case is constant by a factor of around $10$ for the analyzed parameters.
As in Fig.~\ref{Fig:BER_Nmol}-\ref{Fig:BER_a}, Alamouti-type coding does not show any spatial diversity gain compared to a SISO system.

\section{Conclusion\label{Sec:Conclusion}}
In this paper, a diffusion-based molecular $\txt$ MIMO communication system in a 3-D environment was presented.
Channel coefficients were obtained from a trained ANN and incorporated into performance evaluations.
Motivated from the potential of spatial diversity in classical wireless communication,
different spatial diversity algorithms were introduced to the area of MC and their performances were analyzed.
In detail, Alamouti-type coding and repetition MIMO coding were proposed at the transmitter side.
At the receiver side, equal-gain combining that is equivalent to MRC in symmetrical scenarios, was presented as a receiver combining strategy.
In addition, adaptive threshold detection and maximum-likelihood sequence estimation were adapted to the $\txt$ MIMO scenario. The diversity gain was analyzed by numerical simulations. We leave practical algorithms for unsymmetrical cases for our future work. 
%

\section*{Acknowledgment}
The work of H. B. Yilmaz and C.-B. was in part supported by the Basic Science Research Program (2017R1A1A1A05001439) through the NRF of Korea. 
\bibliographystyle{IEEEtran}
\bibliography{mybibs_MIMO_diversity}

\end{document}